\documentclass[aps,twocolumn,prl,color,psfig,epsf,nobalancelastpage]{revtex4}
\usepackage{amsmath}
\usepackage{color}
\usepackage{caption}
\usepackage{amsfonts}
\usepackage{epsf}
\usepackage{graphicx}
\usepackage{epstopdf}
\usepackage{ulem}
\usepackage{balance}
\usepackage{bm}

\begin{document}

\title{Polymer Modelling Predicts Chromosome Reorganisation in Senescence}
\author{Michael Chiang$^1$, Davide Michieletto$^{1,2}$, Chris A. Brackley$^1$, Nattaphon Rattanavirotkul$^2$, Hisham Mohammed$^3$, Davide Marenduzzo$^1$, Tamir Chandra$^2$}
\affiliation{$^1$ SUPA, School of Physics and Astronomy, University of Edinburgh, Peter Guthrie Tait Road, Edinburgh, EH9 3FD, UK, \\
$^2$ MRC Human Genetics Unit, MRC Institute of Genetics and Molecular Medicine, University of Edinburgh,\\ Western General Hospital, Crewe Road South, Edinburgh, EH4 2XU, UK \\
$^3$ Cancer Early Detection Advanced Research Center, Knight Cancer Institute, Department of Molecular \& Medical Genetics,\\Oregon Health \& Science University, Portland, OR 97239, USA}

\begin{abstract}
Lamina-associated domains (LADs) cover a large part of the human genome and are thought to play a major role in shaping the nuclear architectural landscape. Here, we perform polymer simulations, microscopy and mass spectrometry to dissect the roles played by heterochromatin- and lamina-mediated interactions in nuclear organisation. Our model explains the conventional organisation of heterochromatin and euchromatin in growing cells and the pathological organisation found in oncogene-induced senescence and progeria. We show that the experimentally observed changes in the locality of contacts in senescent and progeroid cells can be explained as arising due to phase transitions in the system. Within our simulations LADs are highly stochastic, as in experiments. Our model suggests that, once established, the senescent phenotype should be metastable even if lamina-mediated interactions were reinstated. Overall, our simulations uncover a generic physical mechanism that can regulate heterochromatin segregation and LAD formation in a wide range of mammalian nuclei.
\end{abstract}

\maketitle

\section{Introduction}
The spatial organisation of interphase chromosomes in metazoans is characterised by folding into a hierarchy of structures, from ``topologically associated domains'' (TADs)~\citep{Bonev2016} to compartments~\citep{Rao2014} and chromosome territories~\citep{Cremer2001}. TADs are currently understood as originating from the action of processive~\citep{Fudenberg2016} or diffusing~\citep{Brackley2017prl} cohesin complexes, whereas the establishment of segregated active and inactive genomic compartments is naturally explained by the polymer-polymer phase separation of chromatin segments bearing similar epigenetic marks (i.e., post-translational modifications of histone proteins)~\citep{Jost2014,Brackley2016nar,Brackley2017biophysj,Pereira2018,Erdel2018}. At the scale of chromosomes, their territorial nature may be explained by the slow dynamics of chromatin during interphase~\citep{Rosa2008,Kang2015}. Yet, the large-scale nuclear organisation displays a further level of segregation which is less well understood: one in which heterochromatin (HC) is preferentially found in specific concentric layers either near the nuclear lamina (NL) or the nucleolus~\citep{VanSteensel2017,Solovei2016,Cook2009}, while euchromatin (EC) is enriched in the interior, or middle, layer~\citep{Solovei2016}. Regions of the genome that are preferentially bound to the NL, so-called lamina-associated domains (LADs), are strongly enriched in long interspersed nuclear elements (LINEs) and are associated with gene repression~\citep{VanSteensel2017}. Intriguingly, LADs display a substantial overlap with nucleolus-associated chromatin domains (NADs)~\citep{Nemeth2010}, and together they cover more than a third of the human genome. 

A common approach to analysing lamina-mediated nuclear organisation is through perturbation studies in which the concentric layering of HC and EC is disrupted. An important example, on which we focus here, is that of cellular senescence. A popular way to trigger senescence is to expose cells to stress -- e.g., mitotic stresses or DNA damage. In this way, the cell cycle can be permanently arrested within days~\citep{Chandra2016}. Such cells typically harbour large HC bodies known as senescence-associated heterochromatin foci (SAHF)~\citep{Narita2003,Chandra2016}. This nuclear phenotype, which is a hallmark of stress-induced senescence, can be visualised by DNA stains. It is also possible to study cellular senescence by isolating cells from prematurely ageing (progeroid) patients, which harbour a mutation in the lamin A/C (\emph{LMNA}) gene. In both stress-induced senescence and progeria, there is a weakening of the nuclear lamina and of lamina-chromatin interactions; however, qualitative differences between the two states have been reported for markers of HC. In progeria (e.g., in Hutchinson-Gilford Progeria  Syndrome, HGPS), there is a reduction in the HC mark histone 3 lysine 9 tri-methylation (H3K9me3), whereas stress-induced senescence seems to be associated with an increase in some HC associated proteins, such as heterochromatin protein 1 (HP1) and Core histone macro-H2A (mH2A), but not in the H3K9me3 mark. Additionally, progeroid cells are devoid of the SAHF found in stress-induced senescence.

Although SAHF were first identified more than a decade ago, the connection between the changes in HC proteins and SAHF formation has not been resolved. We still have no clear understanding of the function of SAHF, with contradicting suggestions ranging from pro-proliferative activity to irreversible seals of the senescence arrest. In an effort to gain insight into the role of these two key players, lamina and HC, in the nuclear dynamics in cellular senescence, we developed and studied a model based on concepts from polymer physics. Our model focused on HC-mediated and NL-mediated interactions with chromatin. We analysed chromatin immunoprecipitation with sequencing (ChIP-seq) data~\citep{Chandra2012,Sadaie2013} to accurately capture chromosome-NL interactions in human cells. Surprisingly, by varying only two parameters -- the strength of HC-HC and HC-NL interactions -- our simulations predicted a range of distinct nuclear architectures that are in good qualitative agreement with the known large-scale genome organisation in growing, senescent and progeroid cells. We specifically performed fluorescence microscopy to show that our simulations yield chromatin organisation consistent with that in senescent cells. We also generated chromatin mass spectrometry data for senescence to obtain a more quantitative understanding of the changes in HC markers. More importantly, we found that our model can quantitatively recapitulate the change in the network of chromatin contacts in senescence and progeria observed in Hi-C experiments~\citep{Mccord2013,Chandra2015}. Our simulations show that distal contacts are enriched in senescent cells with respect to growing cells, while they are depleted in progeroid cells.

Within our model, we found that LADs are stochastic and display cell-to-cell heterogeneity, as have been observed experimentally~\citep{Kind2013,Kind2015}. We also used the model to analyse the growing-senescence transition. Our simulations predicted that this transition should be abrupt, so the senescent state should be metastable even when NL-chromatin interactions are partially re-established. This suggests a biophysical reason for the observation that senescent cells with SAHF do not re-enter the cell cycle. Our model also predicted that the dynamics of LADs detachment from the NL upon entering senescence should follow the kinetics observed in polymer desorption~\citep{Johnson1992,Douglas1993}.

Our results demonstrate that polymer physics principles can explain the concentric organisation of HC and EC in growing, senescent and progeroid human cells. At the same time, because our model is developed from first principles -- i.e., no data fitting from chromosomal contact maps -- we can readily export it to study the organisation in other cell types or other organisms. For instance, the inverted organisation found in the nuclei of rod cells of nocturnal mammals~\citep{Solovei2009,Solovei2016} entails the desorption of LADs from the lamina and the formation of a dense HC core. It is likely that the principles underlying this organisation are similar to those we discuss here for stress-induced senescence. 

\section{Results}

\subsection{A Model for Interphase Chromosomes that Incorporates Lamina-Mediated Interactions}
Most previous polymer models for interphase chromosomes have focused on intra- and inter-chromosomal interactions and have largely neglected lamina-associated constraints on chromatin folding~\citep{Rosa2008,Barbieri2012,Jost2014,DiPierro2016}. In contrast, here we developed a polymer model which incorporates lamina-heterochromatin interactions to dissect the effects of lamina tethering on chromosome folding and nuclear positioning. We performed Brownian dynamics simulations which follow the motion of chromatin within a realistic viscous environment and subject to effective potentials modelling steric interactions and protein-mediated attraction~\citep{Brackley2016nar,Barbieri2012,DiPierro2016,Fudenberg2016}.

\paragraph{The Chromosome} 
LADs form large continuous blocks of chromatin which are strongly enriched in HC marks~\citep{Pickersgill2006,Guelen2008,Kind2013,Kind2015}. To account for NL-mediated interactions, we coarse-grained human chromosomes into flexible chains made of beads, each representing 10 kb. A comparison with microscopy data showed that these were consistent with a bead diameter $\sigma \approx 70$ nm (see STAR methods for details).  Each bead is identified as either HC or EC based on the enrichment of H3K9me3~\citep{Chandra2012} and LaminB1~\citep{Sadaie2013} ChIP-seq signal in the corresponding genomic region (see STAR Methods and Figure~\ref{fig:model}A).  The LaminB1 signal has been shown to correlate well with LADs identified from DamID data~\citep{Sadaie2013}. At this resolution of 10 kb per bead, a simulation of the whole genome would require about 600,000 beads, resulting in simulations too long to explore a large parameter space. In addition, the conformations we seek to explore, like SAHF, have been shown to operate on a chromosomal or smaller scale~\citep{Zhang2007}. Thus, to render the simulations more computationally feasible, we considered only one chromosome, human chromosome 20, as it contains large regions of both active and inactive epigenetic marks~\citep{Chandra2012}, and it displays a moderate tendency to be near the nuclear periphery~\citep{Branco2008}. We considered a realistic nuclear chromatin density by performing our simulations within a box of linear size $L= 35 \sigma \approx 2.5$ $\mu$m and with periodic boundaries in the $x$-$y$ directions and confined in the $z$ direction, so as to mimic a small portion of the nucleus near the NL (see Figures~\ref{fig:model}A,B). 

\paragraph{The Lamina}
The NL was modelled by adding a thin layer (about $1\sigma \approx 70$ nm) of randomly positioned beads representing lamin proteins at the top of the simulation box (see Figures~\ref{fig:model}A,B). The interactions between HC and NL are mediated by a variety of proteins ~\citep{Guelen2008,Olins2010,Solovei2013,Poleshko2013,VanSteensel2017}. For example, an anchor protein between NL and HC is the lamin B receptor (LBR), which has been shown to play an important role in forming the peripheral HC~\citep{Solovei2013} and interacting with HP1~\citep{Ye1997,Olins2010}. Evidence for these protein-protein interactions and the fact that LADs are largely comprised of HC~\citep{Pickersgill2006,Guelen2008,Kind2013,Kind2015} were accounted for by setting an effective attraction between HC and NL beads via a phenomenological potential with strength given by the energy $\epsilon_{\text{HL}}$ (see STAR Methods). Further, in oncogene-induced senescence, HC-NL interactions could be negatively affected by an increase in density of nuclear pore complexes~\citep{Boumendil2019}.

HP1 has been shown to dimerise \emph{in vitro} and \emph{in vivo} and has been thought to mediate HC compaction~\citep{Platero1995,Ye1997,Smothers2000}. In line with observations on HP1 dimerisation~\citep{Strom2017} and previous modelling of intra-chromatin folding~\citep{Jost2014,Brackley2016nar,Brackley2016nucleus,Gilbert2017}, we hypothesised that HP1 mediates HC-HC interactions and thus set self-association interaction between HC beads via the same potential as for HC-NL interaction but with a strength given by the energy $\epsilon_{\text{HH}}$. We also incorporated a weak interaction between EC beads (strength $\epsilon_{\text{EE}}=0.2$ $k_BT$) to model in a simple way promoter-enhancer interactions and cohesin-mediated chromatin looping~\citep{Fudenberg2016,Pereira2018}. We did not include any direct EC-HC attraction: this is in line with previous literature~\citep{Brackley2016nar}, as multivalent chromatin-binding proteins tend to bind either HC or EC (hence leading to HC-HC or EC-EC self-association), but rarely bridge HC and EC segments. Our model for HC and EC interactions is similar to, for instance, those of~\citet{Jost2014} and \citet{Falk2019}.

In summary, the behaviour of our model depends on three parameters: $\epsilon_{\text{EE}}$, $\epsilon_{\text{HH}}$ and $\epsilon_{\text{HL}}$. Because we are especially interested in heterochromatin and lamina-mediated interactions, we have fixed $\epsilon_{\text{EE}}$ and have independently varied the two remaining parameters, $\epsilon_{\text{HH}}$ and $\epsilon_{\text{HL}}$, to explore the phase space of chromatin organisation.

\paragraph{The Observables}
To quantitatively characterise the behaviour of our system for different combinations of the parameters $(\epsilon_{\text{HH}},\epsilon_{\text{HL}})$, we computed the distance $\bar{z}$ between the centre of mass of the chromosome and the NL and the average local number density $\rho$ of chromatin beads (see STAR Methods). The first observable $\bar{z}$ quantifies polymer adsorption: a smaller $\bar{z}$ corresponds to a larger degree of adsorption. The second observable $\rho$ quantifies chromatin compactness. It counts the average number of neighbours per unit volume of any chromatin bead. A smaller $\rho$ signifies a more open, extended conformation for the chromatin fibre.

\subsection{Heterochromatin and Lamina Interactions are Sufficient to Capture Chromosomal Conformations Resembling Growing, Senescent and Progeroid Cells}
By performing stochastic polymer simulations across the parameter space $(\epsilon_{\text{HH}},\epsilon_{\text{HL}})$, we discovered that the chromatin fibre exhibits four qualitatively distinct organisations, or phases, corresponding to the four possible combinations of adsorbed/desorbed states (associated with low/high values of $\bar{z}$) and extended/collapsed polymer conformations (associated with low/high values of $\rho$, see Figures~\ref{fig:phase_diagram}A-D). Note that we use the term ``phase'' interchangeably with ``state'' here to refer to the different polymer structures found thermodynamically. This should not be confused with ``phase separation'', a phenomenon which occurs between HC and EC beads in our model when $\epsilon_{\text{HH}}$ is sufficiently strong.

Three of the four phases display morphologies reminiscent of those of healthy (growing), senescent or progeroid mammalian cells (see cartoons in Figure~\ref{fig:phase_diagram}E). 
Specifically, there are two phases with small $\bar{z}$: the adsorbed-collapsed (AC) phase with $\rho > 0$ and the adsorbed-extended (AE) phase with $\rho \simeq 0$. These phases display a layer of HC adsorbed to the NL and are qualitatively similar to healthy, growing cells. However, the AE phase displays a significant amount of HC intermixed with EC, which is not observed in conventional mammalian cells~\citep{Wu2005,Chandra2012}. Therefore, we identified the AC phase as the closest representative of a cell in the growing state. The two phases with large $\bar{z}$, where no chromatin is adsorbed to the NL, are the desorbed-collapsed (DC) phase with $\rho > 0$ and the desorbed-extended (DE) phase with $\rho \simeq 0$. They display features akin to those observed in senescent and progeroid cells. Particularly, the chromosome structure in the DC phase markedly resembles the nucleus of a senescent cell as observed in our fluorescence microscopy in which HC self-associates into large, SAHF-like bodies surrounded by a corona of EC~\citep{Chandra2012,Chandra2015} (Figures~\ref{fig:phase_diagram}F-J). The DE phase instead displays phenotypes reported in progeroid cell nuclei, including the loss of peripheral HC~\citep{Goldman2004,Shumaker2006,Scaffidi2005} and a large degree of mixing of chromatin regions with active and inactive epigenetic marks~\citep{Mccord2013}. 

Within our simulations, the DC/senescent phase is associated with extensive HC-HC interactions and weak HC-NL interactions. This is in line with previous qualitative observations of the upregulation of HP1 in senescence~\citep{Narita2003}. More quantitatively, we have also assessed the changes in proteins able to mediate HC-HC interactions. We performed chromatin fractionation for growing and senescent cells followed by mass spectrometry (see Figure S1, Table S1 and STAR Methods). We found a consistent upregulation of HP1 proteins and macroH2A, which has been previously implicated in SAHF formation and heterochromatin compaction~\citep{Zhang2005}. These data are therefore consistent with the simulation result that SAHF formation might be driven through HC-HC interactions. Additionally, our mass spectrometry data show that high mobility group box (HMGB) proteins are downregulated in senescent cells, similarly to what was previously reported~\citep{Zirkel2018}. 

Given the low-dimensionality of the parameter space scanned by our simulations, it is remarkable that our model can capture key qualitative features of chromosomal organisation in a range of different cell states. In light of this, we concluded that the two key ingredients of our model, HC-HC and HC-NL interactions, must be the major driving forces of chromatin folding in these cell states and can guide the dynamical reorganisation of the nuclear architecture upon transitions to different physiological and pathological conditions. 

\subsection{Simulations Reproduce Changes in the Network of Chromatin Contacts Observed in Senescence and Progeria}
Hi-C experiments have shown that the pattern of chromatin contacts differs largely when comparing growing cells to senescent and progeroid cells~\citep{Mccord2013,Chandra2015}. In particular, it has been demonstrated that distal contacts are more abundant in senescent cells than in growing cells. The reason for this remains elusive, and elucidating it is a key aim of our current work. We therefore used our simulations to quantitatively address how chromatin folding changes in the growing (AC), senescent (DC) and progeroid (DE) phases in our phase diagram.

Hi-C-like contact maps generated from our model show that while senescent simulations display more long-range contacts with respect to growing simulations, those for progeria show a lack of distal contacts. Strikingly, this is the same qualitative behaviour that has been observed in experimental contact maps for growing, senescent and progeroid cells~\citep{Chandra2015,Mccord2013} (Figures~\ref{fig:oci}A,B).

To characterise the change in the network of contacts quantitatively, we calculated the ratio of distal to local contacts at each chromosomal region, also known as the ``open chromatin index'' (OCI)~\citep{Chandra2015} (Figures~\ref{fig:oci}C-G). We set the threshold separating local from distal contacts at 2 Mb (which is roughly the upper limit for the size of a TAD~\citep{Dekker2015}, see STAR Methods). We also found that variations in the threshold do not lead to qualitative change in our conclusions (Figures S2A,B). The difference in OCI, $\Delta$OCI, quantifies the changes in the network structure upon transitioning from the growing to the senescent (or progeroid) phase.

In our simulations, we could detect a dramatic change in OCI in the senescent (DC) and progeroid (DE) phases compared to the growing (AC) phase. First, in simulations, the transition from the growing (AC) to senescent (DC) phase is characterised by an overall positive and statistically significant $\Delta$OCI (two-sample Kolmogorov-Smirnov (KS) test: $D = 0.25$, $p < 10^{-4}$; a larger $D$ means that the two samples are drawn from more separated distributions; Figure~\ref{fig:oci}D). In experiments, $\Delta$OCI for chromosome 20 (excluding inter-chromosomal interactions) is also statistically significant  ($D = 0.50$, $p < 10^{-4}$; Figure~\ref{fig:oci}D). This finding implies that, as qualitatively shown by the contact maps, there is a substantial increase in distal contacts in both experiments and simulations. To quantify the agreement between simulation and experiment, we calculated the Pearson correlation coefficient for the OCI in each state and found $r = 0.53$ ($p < 10^{-4}$) for growing and $r = 0.63$ ($p < 10^{-4}$) for senescence.

Second, the transition from the growing (AC) to progeroid (DC) phase is characterised by a strongly negative $\Delta$OCI, both in simulations ($D = 0.60$, $p < 10^{-4}$) and in experiments ($D = 0.89$, $p < 10^{-4}$; Figure~\ref{fig:oci}E). Here the correlation in the OCI between simulation and experiment is $r = 0.63$ ($p < 10^{-4}$) for growing and $r = 0.48$ ($p < 10^{-4}$) for progeria.
 
It is remarkable that while both senescence and progeria cell states display a global loss of chromatin-lamina interactions, they show an opposite change in chromatin contacts captured by the OCI. We reasoned that this difference may be associated with the appearance of segregated HC compartments in senescence but not in progeria. To verify this hypothesis, we constructed scatter plots showing the OCI value of each bead, colour-coded based on its chromatin (EC/HC) state (Figures~\ref{fig:oci}F,G). These plots reveal that the change in OCI associated with the growing-senescence transition is stronger in HC-rich regions compared with EC-rich regions (blue points appear further from the diagonal). This trend is less substantial when comparing growing with progeroid cells. This result is consistent with the finding that GC-poor isochores exhibit a larger change in their interaction network between growing and senescence~\citep{Chandra2015}, and it further illustrates the fundamental role played by the competition between HC-HC and HC-NL interactions in shaping the nuclear landscape and its reorganisation upon changes in physiological conditions. Furthermore, this finding constitutes compelling evidence that SAHF, present in senescent cells but notably absent in progeria, may be mediating the emergence of long-range chromatin contacts by forming a polymer-polymer phase separation between EC and HC~\citep{Brackley2013,Brackley2016nar,Michieletto2016,Erdel2018}. 

To rationalise the observed opposite change in distal contacts in senescent and progeroid cells, we considered the decay of the contact probability $P(s)$ of two chromatin segments as a function of their genomic distance $s$, which can be extracted from Hi-C maps~\citep{Barbieri2013}. Classic results from polymer physics predict that the contact probability should scale as $P(s) \sim s^{-c}$, where $c$ is the contact exponent; different polymer organisations are associated with different values for the exponent~\citep{DeGennes1979,Rubinstein2003,Mirny2011}.

Motivated by the distinct organisations found in our simulations for the growing (AC), senescent (DC) and progeroid (DE) phases, we predicted that contact exponents should differ for these phases. Measuring $c$ from our contact maps confirmed this hypothesis (see Figure S3). In particular, we found the relation $c_{\text{DC}} < c_{\text{AC}} < c_{\text{DE}}$. This inequality is consistent with the change in distal contacts. Because contact probabilities are normalised, a smaller value of $c$ leads to a shallower decay in the contact probability, hence a shift favouring non-local over local contacts. Therefore, as $c_{\text{DC}} < c_{\text{AC}}$, the decay in contact probability in the senescent (DC) phase is shallower with respect to the growing (AC) case, thus non-local contacts are {\it more} likely. Instead, as $c_{\text{DE}} > c_{\text{AC}}$, the contact probability decays more rapidly in the progeroid (DE) phase, hence non-local contacts are now {\it less} likely. Given that our model requires minimal inputs from biological data and is largely based on the polymeric nature of chromatin, our results show that the observed change in contact patterns between different cell states can be attributed to a change in the physical organisation of the chromatin fibre.

\subsection{Simulations Reproduce Experimental Observations of Cell-to-Cell Variability in LADs}

\begin{figure*}[ht]
\begin{center}
\includegraphics[width=0.9\columnwidth]{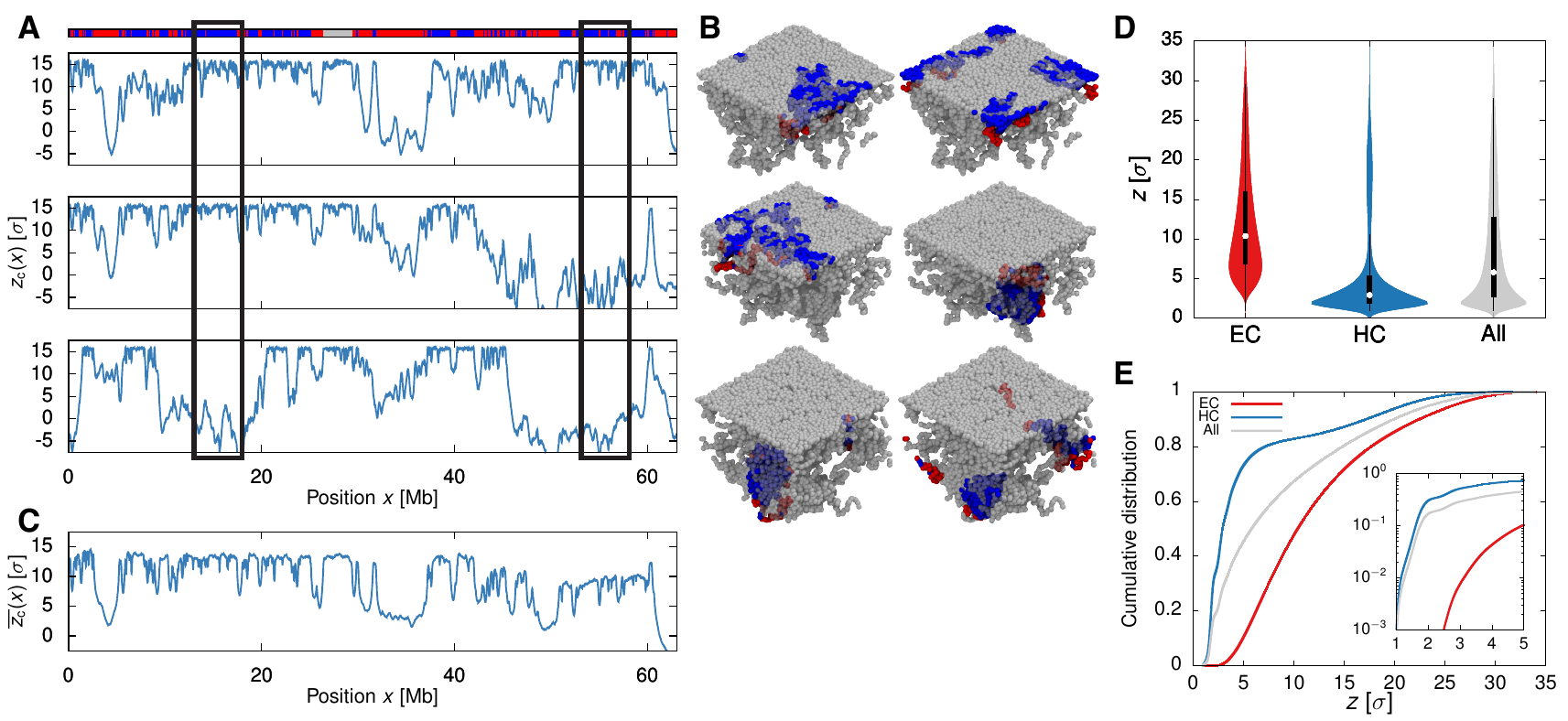}
\end{center}
  \caption{{\bf{Simulations show stochasticity in chromatin contacts with the NL.}} (A) Plots showing the distance $z_c$ of each chromatin bead from the horizontal plane at the centre of the simulation box for three simulation runs in the growing (AC) phase ($\epsilon_{\text{HH}} = 1.0, \epsilon_{\text{HL}} = 1.6$). As the NL is located at the top of the box $(z_c \approx 17\sigma)$, a high value in $z_c$ indicates the bead is close to the NL. The $x$-axis shows the genomic position corresponding to each bead. The top track indicates the chromatin state along the polymer. (B) Snapshots of the simulation runs colouring only the beads within the two highlighted regions in (A) (rectangular boxes). These figures reveal that the same chromatin segment can reside in different positions relative to the NL in different runs, which is consistent with experimental observations that LADs associate with the NL stochastically. (C) The average value of $z_c$ for each bead over 20 simulations. (D) Violin plots showing the distributions of the distance $z$ from the NL for EC, HC and all beads. (E) Cumulative distributions of $z$ for EC, HC and all beads. The inset shows the distributions (in log-linear form) for small $z$.}
  \label{fig:wall_contact}
\end{figure*}

Lamina association is a major regulator of nuclear architecture, and LADs cover more than 30\% of the human genome~\citep{VanSteensel2017}. Yet, each cell has only a limited amount of surface that is available for chromatin to interact with. For this reason it has been conjectured, and then shown experimentally, that LADs display cell-to-cell variability, appear stochastically and are not conserved in daughter cells~\citep{Kind2013,Kind2015}. By measuring the adsorption of beads onto the NL in single, independent polymer simulations, which we associated with individual cells, we found that this variability is captured by our model (Figure~\ref{fig:wall_contact}A), due to the stochastic nature of our Brownian dynamics simulations. 
Thus, even though all HC beads can in principle be adsorbed to the NL, we found that the same HC segment can be adsorbed in one simulation, but not in another (Figure~\ref{fig:wall_contact}B). In other words, after averaging over all simulation replicas (equivalent to averaging over a population of cells), we found that the average distance from the wall for each bead differs noticeably from that in single simulation runs (Figure~\ref{fig:wall_contact}C). This result emphasises that population averaged information on chromatin conformation, such as Hi-C, may not fully reflect the conformation assumed in single cells~\citep{Nagano2015}. It is also consistent with recent work which combines Hi-C and high throughput fluorescence \textit{in situ} hybridization experiments to show that spatial genome organisation is highly heterogeneous at the single cell level~\citep{Finn2019}. 

We also noted that some EC beads were close to the NL, even though there was no explicit attractive interaction between the two in the model. This is due to the chromatin context in which a given EC bead is embedded -- i.e., a EC-rich segment within a large HC-rich chromosomal region is likely to be ``dragged along'' and co-adsorbed onto the NL (see Figures~\ref{fig:wall_contact}B,D-E). This observation indicates the importance of considering the epigenetic context and polymeric nature of neighbouring chromatin when determining the spatial location and function of a specific locus. 

\subsection{A Sharp Phase Transition Separates Growing and Senescent States}

After showing that our polymer model can capture many complex features observed in growing, senescent and progeroid cells, we employed it to obtain further insight into the nature of the transition between different states of a cell. This has biological relevance, as the nature of the transition determines the stability of the growing and senescent phases upon external perturbations, such as a sudden change in the number of active HP1 or lamin proteins.

To characterise the nature of the phase transition between the growing and the senescent phase, we asked whether this transition is associated with hysteresis~\citep{Michieletto2016}. To this end, we recorded the instantaneous value for the distance $\bar{z}$ between the chromosome's centre of mass and the NL as we slowly varied the strength of HC-NL interaction, $\epsilon_{\text{HL}}$, between two values known to be deep into the respective phases (Figures~\ref{fig:hysteresis}A-B). By starting from a large value for $\epsilon_{\text{HL}}$ (AC, or growing, phase), slowly reducing it to a low value (DC, or senescent, phase), before reversing the process and slowly increasing it again, we found that the change in $\bar{z}$ does not follow the same pattern in each direction. That is to say, the phase transition does not occur at the same critical value of HC-NL interaction strength, and the system is thus history-dependent. We also observed that close to the transition, the distribution of distances between the chromosome and the lamina is bimodal, signifying that the system is bistable (i.e., for the same parameter value, it can either be desorbed or adsorbed; Figure~\ref{fig:hysteresis}C).

A hysteresis cycle and bistability are well-known hallmarks of a sharp (typically first-order) phase transition, where each of the two states are metastable in a range of parameter values~\citep{Binder1987}. Biologically, the fact that the senescent state is metastable is important, as it suggests that the passage from growing to senescence is likely irreversible, or at least that it can be reverted only by energy expenditure (as a strong perturbation is required to drive the senescent state out of its local free energy minimum). A possible reason for this phenomenology is that the EC-rich chromosomal regions surrounding the SAHF in the senescent (DC) phase (Figure~\ref{fig:phase_diagram}) provide a large entropic barrier that needs to be overcome in order for the HC beads to be adsorbed to the NL~\citep{Johner1990}. In other words, even if the HC beads within the SAHF can bind the NL, a large-scale rearrangement of the surrounding halo of EC beads is required for the HC and NL beads to come into contact. This stabilises the DC phase. A similar argument suggests that the growing (AC) phase is also stable to perturbations as the transition into the DC phase entails a large-scale rearrangement to form the SAHF surrounded by the EC corona, and this rearrangement is likely associated with another free energy barrier.

LADs detachment from the nuclear periphery is a common phenotype in cellular senescence; however, the dynamics of such process is not well understood. Here, we employed our model to provide a prediction on how LADs separate from the lamina as cells enter senescence (for instance, as a response to DNA damage). Specifically, we monitored the distance of chromatin segments from the NL and the fraction of adsorbed segments over time after the HC-NL interaction, $\epsilon_{\text{HL}}$, was suddenly reduced, mimicking the loss of lamina interactions at the onset of senescence (Figures~\ref{fig:hysteresis}D-G). We observed that the decay in the fraction of adsorbed segments is non-exponential. Previous polymer studies have suggested that such a decay trend can arise if the desorption process is rate-limited by the diffusion of chain segments away from the surface~\citep{Johnson1992,Douglas1993}. If future experiments following the separation of LADs from the NL corroborate this trend, it would suggest that polymer dynamics of chromatin play a considerable role in the time scale of forming senescence-associated phenotypes (such as SAHF).

\section{Discussion}

In this work, we proposed and studied a polymer model for nuclear organisation which explicitly takes into account chromatin-lamina interactions. A key result is that by modifying only two parameters -- the HC-HC self-interaction and the HC-lamina (HC-NL) interaction -- our model displays four possible distinct phases, three of which recapitulate nuclear architectures that are biologically relevant. The adsorbed-collapsed (AC) phase corresponds to the morphology of conventional growing cells, where HC forms a layer close to the lamina, and HC-HC interactions drive (micro)phase separation of HC and EC compartments. In stress-induced cellular senescence and progeria, chromatin-lamina interactions are disrupted; hence their corresponding nuclear structure is captured by desorbed phases, where HC-NL interactions are weak. More specifically, the phenotype of cellular senescence, which is associated with extensive HC foci, corresponds to our desorbed-collapsed (DC) phase, where HC-mediated interactions still drive clustering (HC-EC phase separation) but HC is not bound to the lamina. Instead, the desorbed-extended (DE) phase is consistent with the sparse chromosome organisation of progeroid cells.

This classification is in line with qualitative expectations from existing biological models, but the link we make here to polymer physics allows us to further quantitatively explain additional features of nuclear organisation in growing and senescent cells which were previously mysterious. An important result is that, if we view the stress-induced entry to senescence as a phase transition (between the AC and DC phases), then our simulations reproduce the experimental observation -- found by Hi-C -- that intra-chromosomal contacts should become less local and longer-range in senescence~\citep{Chandra2015}. In contrast, when a growing cell becomes progeroid, simulations predict that contacts should become more {\it local}. Notably, by re-analysing Hi-C data for HGPS cells~\citep{Mccord2013}, we found that similar trends occur {\it in vivo}. Theories of polymer looping predict that the probability of contact decays as a power law of the genomic distance (i.e., distance along the backbone). Measuring the power law exponent in simulations, we found it to differ in the different states of the cell (i.e., healthy, senescent and progeroid). In line with the change in contact patterns, the steepest decay (favouring local contacts) was found in the progeroid phase, and the shallowest one (favouring distal contacts) in the senescence phase.

The finding that the inter-chromatin contact network changes in opposing ways in progeroid and senescent cells might seem at first surprising, as progeria is normally viewed as the first step towards senescence~\citep{Chandra2015}. However, our model shows that this result is a consequence of the fact that growing, senescent and progeroid architecture correspond to different thermodynamic phases for interphase chromosomes in the nucleus. 

Our polymer physics framework also yielded additional predictions: some of these conform well with existing experimental evidence, while others could be tested by new experiments in the future. For example, within our model lamina adsorption is stochastic: different HC domains bind the lamina in different simulations, and not all HC is absorbed. This is in line with single-cell DamID experiments which show that not all LADs contact the lamina in every cell, and that the selection of which LADs are bound to the nuclear periphery in any given cell is largely random~\citep{VanSteensel2017}. In addition, our model predicted that the transition between growing and senescent cells should be sharp and associated with memory (or hysteresis), so that each state, once established, should be metastable even in parameter space regions where it is normally unstable. This provides an appealing mechanism to explain the remarkable stability of senescent cells: once SAHF form, the cell virtually reaches a thermodynamic dead end (i.e., it is unlikely to ever re-enter the normal cell cycle of proliferating cells, to grow and divide again). This permanent growth arrest may be a mechanism by which cells can prevent stresses, e.g., induced by an oncogene or by DNA damage, from inflicting further harm to the organism.

As for predictions which could be tested in the future, we suggest that these would entail studying the dynamics of chromosome desorption following the entry into senescence. First, we expect that SAHF should form by coarsening, and the associated growth laws could be compared between simulations and live cell microscopy. Second, we suggest that monitoring the amount of adsorbed chromatin as a function of time will be of interest: this will test our predictions that its decay is non-exponential, and may uncover spatiotemporal correlations between LAD dynamics. It would also be instructive to assess the role of the confining geometry in determining the final morphology of the system. In stress-induced senescence, desorption leads to the formation of a SAHF for each chromosome. In the rod cells of nocturnal mammals, HC also detaches from the lamina but forms a single, larger aggregate~\citep{Falk2019}. It is attractive to think that this may be due to the smaller size of retinal cells with respect to senescent cells: does the enhanced confinement push HC into larger clusters? It might be possible to perturb the geometry of senescent cells to further address the role of nuclear geometry in chromosome architecture: this question could be asked both experimentally and in simulations. Finally, our work shows that the phase separation of HC and EC couples to nuclear topography and architecture in a profound way, and it would be important to find out whether the phase-separated and layered organisation in growing cells offers any functional advantage with respect to the organisation in other cell states. 

\section{Acknowledgements}
D. Mi., C. B. and D. Ma. acknowledge the European Research Council for funding (Consolidator Grant THREEDCELLPHYSICS, Ref. 648050). M. C. acknowledges the Carnegie Trust for the Universities of Scotland for PhD studentship funding.  T. C. is supported by a Chancellor's Fellowship held at the University of Edinburgh and the MRC Human Genetics Unit. N. R. was supported by a PhD studentship funded by the Wellcome Trust Sanger Institute and the Royal Thai Government. We would like to acknowledge Cambridge Centre for Proteomics for mass spectrometry analysis. We would also like to thank N. Robertson for preparing the bioinformatics data required for the simulations. 

\section{Author Contributions}
M. C., D. Mi., C. B., D. Ma. and T. C. designed research and discussed results. M. C. performed simulations. M. C. and C. B. analysed Hi-C data. T. C., H. M. and N. R. performed imaging, chromatin fractionation and mass spectrometry experiments. All authors wrote the paper.

\section{Declaration of Interests}
The authors declare no competing interests. 
\balance

\newpage
\begin{figure*}[ht]
\begin{center}
\includegraphics[width=0.9\columnwidth]{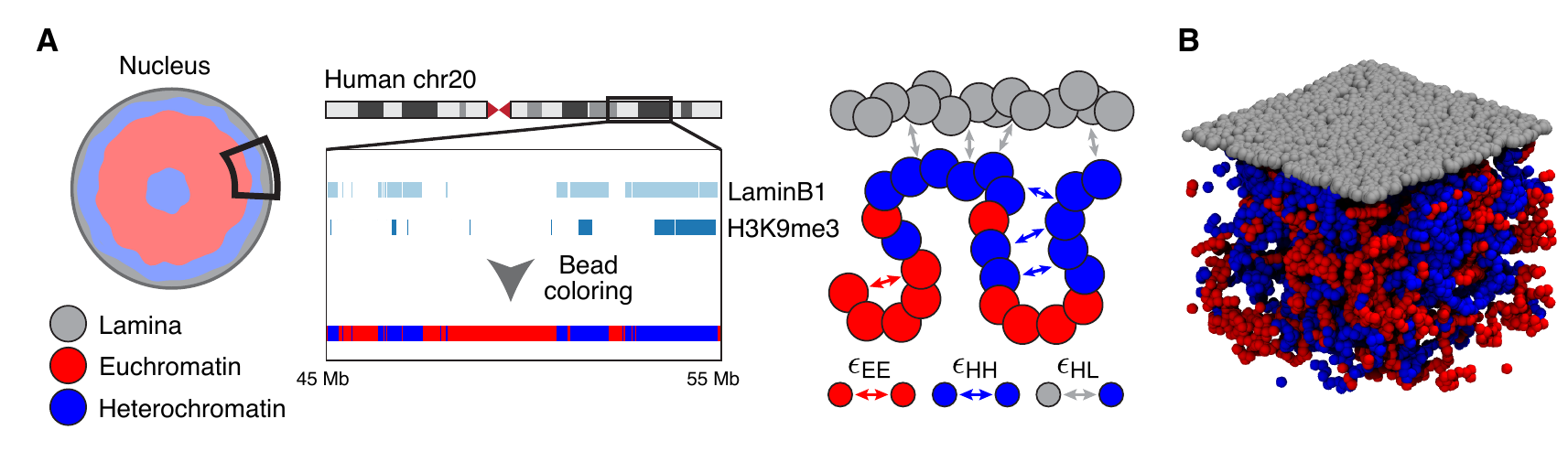}
\end{center}
  \caption{{\bf{A polymer model for lamina-mediated chromosome organisation in different cell states.}} (A) A subsection of the nuclear periphery and human chromosome 20 were simulated. Chromatin was modelled as a flexible bead-spring chain with red beads representing euchromatin (EC) and blue beads representing heterochromatin (HC). The nuclear lamina (NL) was represented as a layer of static beads (grey). Chromatin beads were labelled as HC if the corresponding genomic region is enriched in H3K9me3 and/or LaminB1. Beads corresponding to the centromeric region (26.4 to 29.4 Mb) were also treated as HC. All other beads were labelled as EC. EC and HC beads can interact with beads of the same kind with interaction strength $\epsilon_{\text{EE}}$ and $\epsilon_{\text{HH}}$, respectively. HC beads can also interact with the NL beads with interaction strength $\epsilon_{\text{HL}}$. (B) A simulation snapshot of the model when the HC-HC and HC-NL interactions are weak.}
  \label{fig:model}
\end{figure*}
\begin{figure*}[ht]
\begin{center}
\includegraphics[width=0.9\columnwidth]{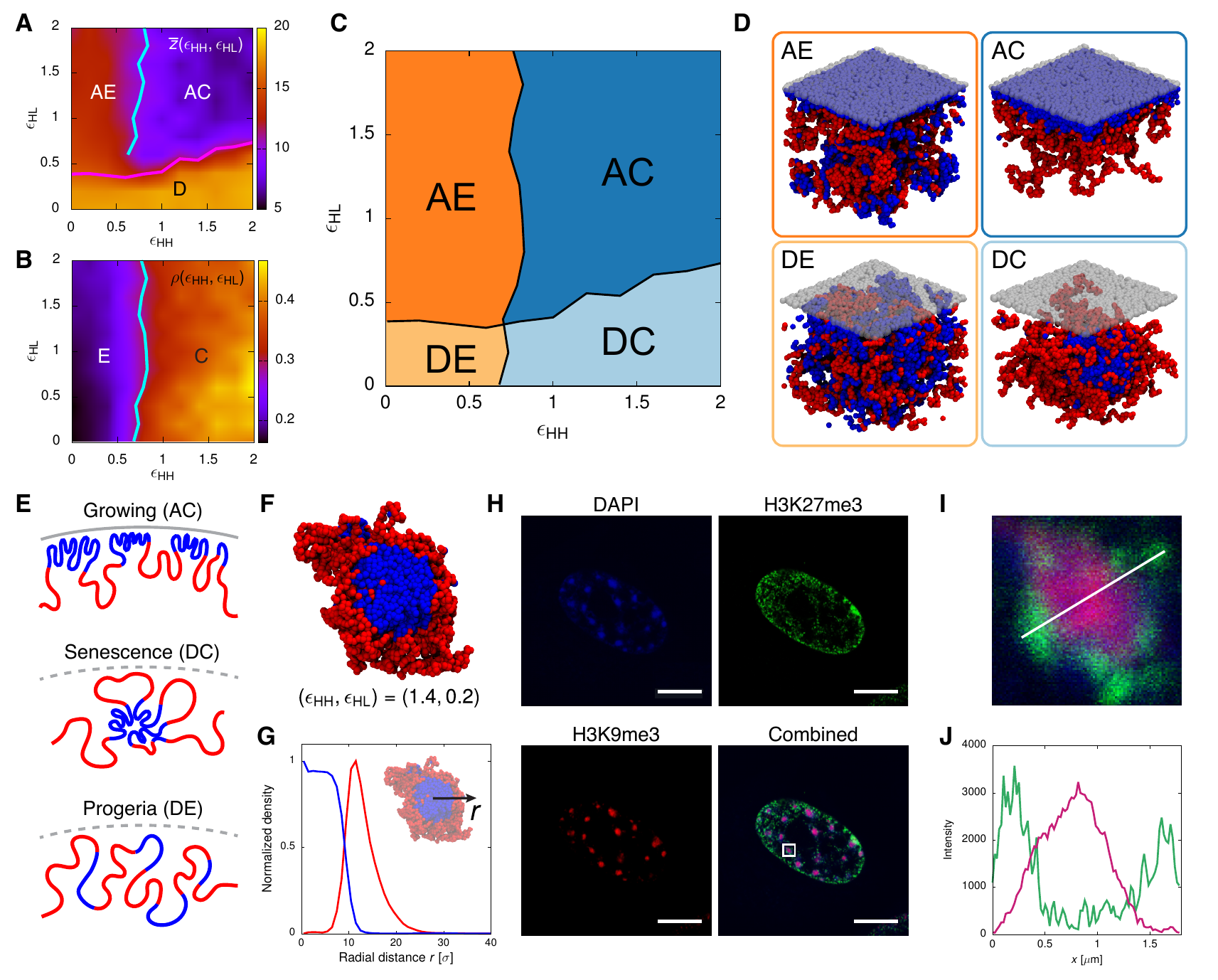}
\end{center}
  \caption{{\bf{Variations of the two model parameters reproduce chromatin organisation in growing, senescent and progeroid cells.}} (A) A heat map showing the distance $\bar{z}$ between the centre of mass of the chromosome from the NL across the parameter space $(\epsilon_{\text{HH}},\epsilon_{\text{HL}})$. 10 simulations were performed for every 0.2 increment in both $\epsilon_{\text{HH}}$ and $\epsilon_{\text{HL}}$ direction. Distances are reported in units of the bead size ($\sigma \approx 70$ nm). The pink line separating the adsorbed and desorbed regime was estimated based on the inflection points in $\bar{z}$ when varying $\epsilon_{\text{HL}}$ for fixed $\epsilon_{\text{HH}}$. $\bar{z}$ also captures the transition between the collapsed and extended phase in the adsorbed regime, as a compact fibre would stay closer to the NL. The cyan line reports the inflection points at which this transition occurs.  (B)  A heat map showing the average local number density $\rho$ of chromatin beads across the parameter space. The cyan line separating the extended and collapsed phase was estimated based on the inflection points in $\rho$. Its location is consistent with that estimated from $\bar{z}$. The interaction strengths for extended and collapsed conformation are also in line with recent studies on chromatin structure in yeast~\citep{Socol2019} and \textit{Drosophila}~\citep{Lesage2019}. (C) A full phase diagram of the four observed phases: adsorbed-extended (AE), adsorbed-collapsed (AC), desorbed-extended (DE) and desorbed-collapsed (DC). Boundary lines are those from (A) and (B). (D)  Simulation snapshots of the four phases. (E) Cartoons of chromatin structures for cells in growing, senescent and progeroid conditions. (F) Cross-section view of a simulation snapshot corresponding to the DC/senescent phase. (G) Time-averaged density profiles of HC and EC as a function of distance $r$ from the centre of the globule. (H) Confocal images of chromosomes in senescent cells with DAPI, H3K27me3 and H3K9me3 staining. Scale bars indicate 10 $\mu$m. (I) Zoomed-in view of a SAHF corresponding to the white square in the combined image in (H). (J) Corresponding intensity profiles for H3K9me3 and H3K27me3 along the white line in (I).}
  \label{fig:phase_diagram}
\end{figure*}
\begin{figure*}[ht]
\begin{center}
\includegraphics[width=0.9\columnwidth]{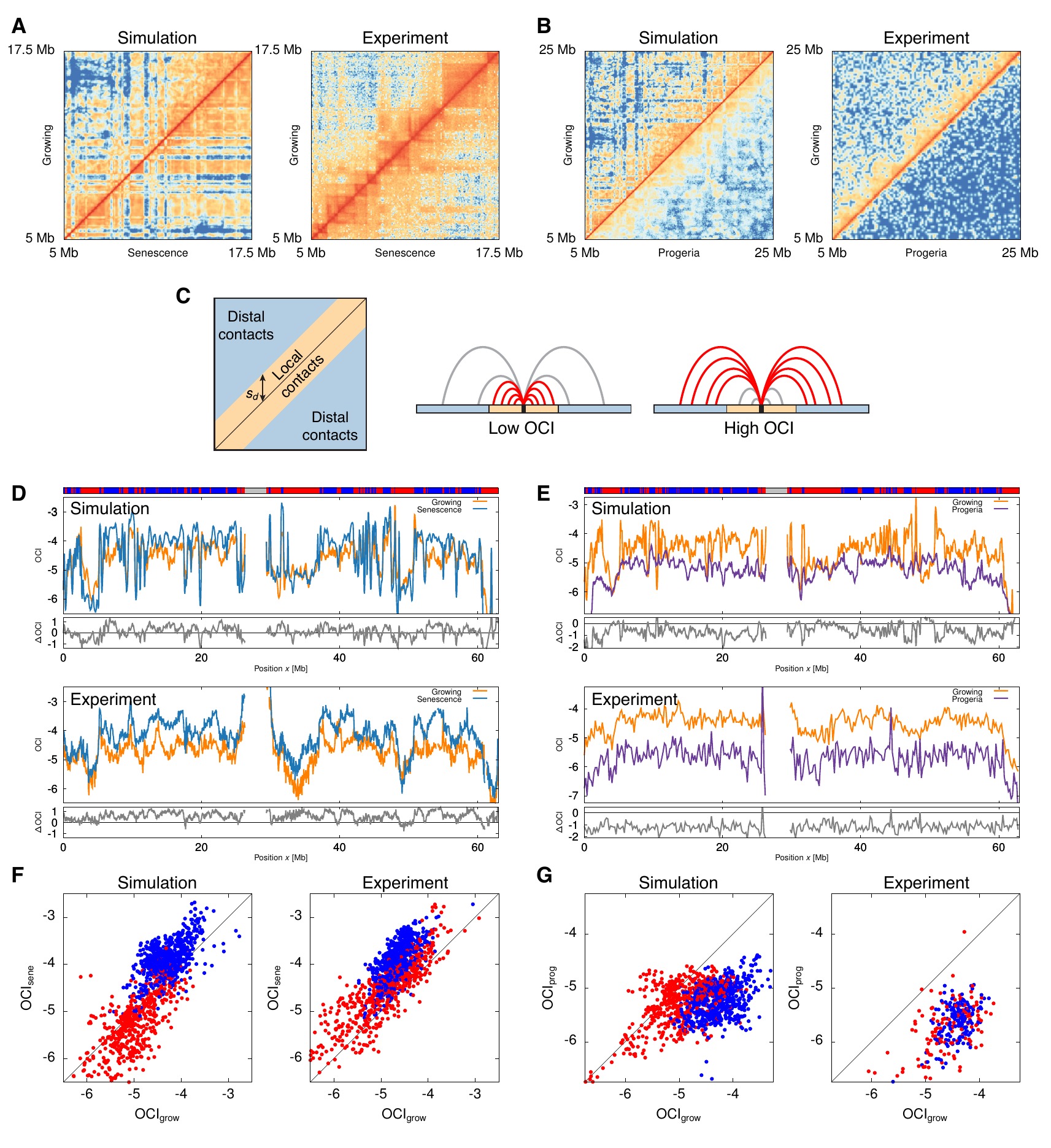}
\end{center}
  \caption{{\bf{Chromatin contact network and the open chromatin index (OCI) in growing (AC), senescent (DC) and progeroid (DE) phase.}}  (A)-(B) Heat maps comparing the contact frequencies between growing and senescence, and between growing and progeria, in simulations and experiments. In our simulations, we used $(\epsilon_{\text{HH}},\epsilon_{\text{HL}}) = (1.0, 1.6)$ for growing, $(1.4, 0.2)$ for senescence and $(0.2, 0.2)$ for progeria. We performed 20 simulations for each case. We used Hi-C data from~\citep{Chandra2015} for the comparison between growing and senescence, and those from~\citep{Mccord2013} for the comparison between growing and progeria. The contact frequencies are plotted in log scale to aid visualisation. (C) A cartoon illustrating the open chromatin index (OCI). Contacts were defined to be local or distal based on a threshold $s_d$, which was set to 2 Mb (close to the maximum size of a TAD~\citep{Dekker2015}; see STAR Methods). (D)-(E) OCI value of each bin in the heat map for growing and senescence, and for growing and progeria. Their difference ($\Delta$OCI) is shown in the bottom track. The top track indicates the chromatin state along the polymer (i.e., red for EC, blue for HC and grey for centromeric region). (F)-(G) Scatter plots of the OCI value of each bin for growing versus senescence, and for growing versus progeria. The colour of each point indicates the chromatin state of the corresponding bin (treating centromeric region as HC).}
  \label{fig:oci}
\end{figure*}
\begin{figure*}[ht]
\begin{center}
\includegraphics[width=0.9\columnwidth]{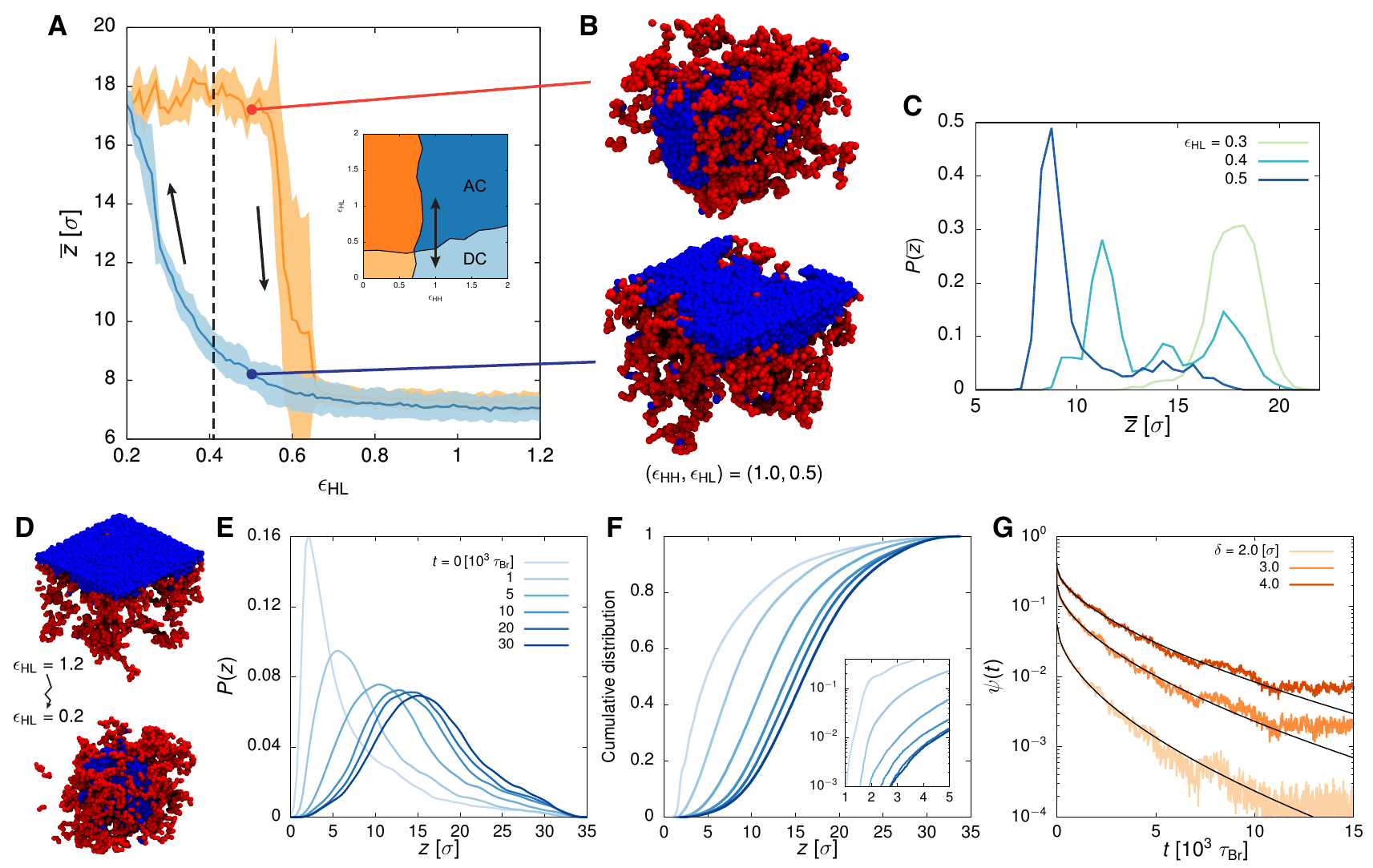}
\end{center}
  \caption{{\bf{The transition between the growing (AC) and senescent (DC) phase is abrupt and history-dependent.}} (A) The distance $\bar{z}$ between the centre of mass of the chromosome and the NL, averaged over five simulation runs in which we fix $\epsilon_{\text{HH}} = 1.0$ and gradually decrease $\epsilon_{\text{HL}}$ from 1.2 to 0.2 over $10^6$ Brownian time steps, before increasing it back to 1.2 over the same time (its path in the phase diagram is shown in the inset). The shaded area around the curve reports the standard deviation of the mean. The dashed line marks the predicted transition point between the adsorbed and desorbed regime at $\epsilon_{\text{HH}} = 1.0$ (see Figures~\ref{fig:phase_diagram}A,C). Hysteresis occurs in the region $\epsilon_{\text{HL}} \sim 0.3$-$0.6$. (B) Snapshots of the system at $\epsilon_{\text{HL}} = 0.5$ showing that it can either be in the AC (growing) or the DC (senescent) phase depending on its history. (C) The probability distribution of $\bar{z}$ at $\epsilon_{\text{HL}} = 0.3$, $0.4$ and $0.5$ ($\epsilon_{\text{HH}} = 1.0$; 50 simulations were sampled for each parameter set). A bimodal behaviour is found when $\epsilon_{\text{HL}} \sim 0.4$, suggesting coexistence of both the AC and DC phase. (D) Snapshots of our simulation modelling the process of LADs detachment from the NL, during which we change $\epsilon_{\text{HL}}$ instantaneously from 1.2 to 0.2. (E) The probability distribution of the distance $z$ of each bead from the NL at time $t$ after $\epsilon_{\text{HL}}$ has been reduced (sampled from 10 simulations). (F) The corresponding cumulative distribution of $z$. The inset shows the distribution for small $z$ in log-linear form. (G) A log-linear plot of the fraction of beads $\psi$ in contact with the NL (those whose distance from the NL is less than $\delta$) at time $t$ after the weakening in $\epsilon_{\text{HL}}$, for three different thresholds of $\delta$. The black curves are stretch exponential fits $f(t) = \kappa\exp\left(-\alpha t^{\beta}\right)$. The fitted stretch exponents $\beta$ are 0.56, 0.58 and 0.61 for $\delta = 2$, $3$ and $4\sigma$, respectively.}
  \label{fig:hysteresis}
\end{figure*}

\clearpage


\section{STAR Methods}

\subsection{Lead Contact and Materials Availability} 
Further information and requests for resources and reagents should be directed to and will be fulfilled by the Lead Contact, Michael Chiang (\url{c.h.m.chiang@sms.ed.ac.uk}). This study did not generate new unique reagents.

\subsection{Experimental Model and Subject Details}
\subsubsection{Cell Lines}
We performed fluorescence microscopy using pLNCX2-ER:ras\textsuperscript{G12V}-expressing IMR90 (plasmid obtained from Addgene \#67844). IMR90 is a normal human foetal lung fibroblast cell line obtained from ATCC. The cells were maintained in DMEM 10\% FCS at 37\textsuperscript{o}C under atmospheric O$_2$ conditions and senescence induced by 4-hydroxytamoxifen (4OHT) in 100nM. Senescence (growth arrest) was triggered after 7 days of 4OHT treatment.

We conducted chromatin fractionation and mass spectrometry analysis using WI-38hTERT/GFP-RAF1-ER, which was a generous gift from Carl Mann. The cells were derived from human embryonic diploid female lung fibroblasts. WI38 cells were transduced with a construct that is composed of EGFP at the N-terminus, the catalytic domain of RAF1 in the middle and the hormone-binding domain of ESR1 at the C-terminus. Protein kinase activity was activated by 4OHT, and hTERT was expressed from a separate locus. The cells were maintained in 10\% fetal bovine serum under 5\% O$_2$ and handled as described in \citet{Jeanblanc2012}. Senescence was induced by 48 hr treatment with 4OHT. 

\subsection{Method Details}
\subsubsection{Chromatin Fractionation and Mass Spectrometry}
\paragraph{Chromatin Isolation}
Cells were trypsinised and resuspended in 10-15 mL DMEM (see Figure S1A,B for a cartoon illustrating the full procedures). Cells were counted, and the same number of cells was collected from each sample by centrifugation and resuspended in cold PBS. Supernatant was removed, and cells were resuspended in no-salt buffer A (DTT, PMSF, Digitonin) and incubated on ice for 10 min. Cells were centrifuged, and supernatant (cytosolic components) was kept for further preparation. Nuclei pellet was resuspended in no-salt buffer A and centrifuged. Supernatant was removed, and nuclei pellet was resuspended in no-salt buffer B (DTT, PMSF) and vortexed occasionally on ice before centrifuging. The nucleoplasmic supernatant fraction was kept for further analysis, and the remaining chromatin pellet was resuspended in no-salt buffer B followed by centrifugation. Supernatant was removed, and chromatin pellet was resuspended in 1$\times$ sample buffer. Suspension was vortexed and heated in boiling water for 5 min to release nuclear-bound proteins.

\paragraph{Chromatin Fractionation}
Cells were harvested and centrifuged. Cell pellets were resuspended in cold lysis buffer and mixed thoroughly. Cell lysate was centrifuged, and supernatant was transferred to a new tube. Protein concentration was determined (BCA Protein Assay Kit). Proteins extracted were reduced, alkylated, digested overnight and labelled as described in the TMT10plex Isobaric Label Reagent Set (see the next paragraph). Labelled samples were analysed by high-resolution Orbitrap LC-MS/MS. Labelled peptides were identified, and reporter ion relative abundance was quantified.

\paragraph{TMT Labelling and Mass Spectrometry}
TMT labelling was performed according to the manufacturer's protocol (\url{https://www.thermofisher.com/order/catalog/product/90110}). Eight samples were digested, and the resulting peptides were labelled with the tags 126, 127C, 128C, 129N, 129C, 130N, 130C and 131, before being combined and lyophilised.

The following LC conditions were used for the fractionation of the TMT samples: desalted peptides were resuspended in 0.1 mL 20 mM ammonium formate (pH 10.0) + 4\% (v/v) acetonitrile. Peptides were loaded onto an Acquity bridged ethyl hybrid C18 UPLC column (Waters; 2.1 mm i.d.~$\times$~150 mm, 1.7 $\mu$m particle size) and profiled with a linear gradient of 5-60\% acetonitrile + 20 mM ammonium formate (pH10.0) over 60 min, at a flow-rate of 0.25 mL/min. Chromatographic performance was monitored by sampling eluate with a diode array detector (Acquity UPLC, Waters) scanning between wavelengths of 200 and 400 nm. Samples were collected in one-minute increments and reduced to dryness by vacuum centrifugation, before being pooled into pairs (in total, 13 paired fractions were generated). Fractions were resuspended in 30 mL of 0.1\% formic acid and pipetted into sample vials.

All LC-MS/MS experiments were performed using a Dionex Ultimate 3000 RSLC nanoUPLC (Thermo Fisher Scientific) system and a QExactive Orbitrap mass spectrometer (Thermo Fisher Scientific). Separation of peptides was performed by reverse-phase chromatography at a flow rate of 300 nL/min and a Thermo Scientific reverse-phase nano Easy-spray column (Thermo Scientific PepMap C18, 2 mm particle size, 100 A pore size, 75 mm i.d.~$\times$~50 cm length). Peptides were loaded onto a pre-column (Thermo Scientific PepMap 100 C18, 5 mm particle size, 100 A pore size, 300 mm i.d.~$\times$~5 mm length) from the Ultimate 3000 autosampler with 0.1\% formic acid for 3 min at a flow rate of 10 mL/min. After this period, the column valve was switched to allow elution of peptides from the pre-column onto the analytical column. Solvent A was water + 0.1\% formic acid and solvent B was 80\% acetonitrile, 20\% water + 0.1\% formic acid. The linear gradient employed was 4-40\% B in 100 min (the total run time including column washing and re-equilibration was 120 min).

The LC eluant was sprayed into the mass spectrometer by means of an Easy-spray source (Thermo Fisher Scientific). All $m/z$ values of eluting ions were measured in an Orbitrap mass analyser, set at a resolution of 70000. Data dependent scans (Top 20) were employed to automatically isolate and generate fragment ions by higher energy collisional dissociation (HCD) in the quadrupole mass analyser, and measurement of the resulting fragment ions was performed in the Orbitrap analyser, set at a resolution of 35000. Peptide ions with charge states of between 2+ and 5+ were selected for fragmentation. The analysis of the mass spectrometry data is discussed in the Quantification and Statistical Analysis section. 

\subsubsection{Fluorescence Microscopy}
IMR90-ER:RAS cells were plated on gelatin-treated coverslips and allowed to attach to the surface overnight. Cells were fixed in 4\% paraformaldehyde (in 1$\times$PBS) for 15 min and washed three times. Fixed cells were permeabilised using 0.2\% Triton X/PBS for 5 min at RT. Primary antibodies (Anti-Histone H3 (tri methyl K9) [Ab8898, Abcam] plus Anti-Histone H3 (tri methyl K27) [Ab6002, Abcam], or (Anti-Histone H3 (tri methyl K36) [Ab9050, Abcam] plus Anti-Histone H3 (tri methyl K9) [05-1242-S, Millipore]) were diluted in 1:1000 ratio (1$\times$PBS) and incubated on the cells for 45 min at RT. The cells were washed with the blocking solution PBS-T (0.1\% Tween in 1$\times$PBS) for 30 min, followed by the incubation with fluorophore conjugated secondary antibodies (Alexa Fluor\textsuperscript{\textregistered} 488 anti-rabbit and Alexa Fluor\textsuperscript{\textregistered} 555 anti-mouse) for 45 min in darkness at RT. After secondary incubation, the cells underwent additional washes with PBS-T for another 30 min before being dried and mounted with Vectashield antifade mounting medium containing 4'-6-diamidine-2-phenyl indole (DAPI). We used the Nikon's A1R point scanning confocal laser-scanning microscope for capturing immunofluorescent images. To optimise the quality of image acquisition, the Nikon's CFI Plan Fluor 60$\times$ oil objective lens was selected to acquire images. The triple band excitation DAPI-FITC-TRITC filter was used to detect fluorescent signals from DAPI, Alexa Fluor\textsuperscript{\textregistered} 488 and Alexa Fluor\textsuperscript{\textregistered} 555, respectively. The laser power was set at 4.00 and the detector sensitivity fixed at 80 for every fluorescence channel. The pinhole was set at 1.2 AU and the images acquired were 2048 px $\times$~2048 px.
 
\subsubsection{Simulation Details}
\paragraph{Computational Modelling of Chromatin and Lamina}
We modelled the chromosome as a flexible bead-spring chain of $N$ beads. Each bead represents a 10 kb chromatin segment, which is equivalent to about 50 nucleosomes. Note that a bead here simply represents a ``blob'' of chromatin, as commonly considered in scaling approaches in polymer physics~\citep{DeGennes1979}, and the organisation of chromatin within a bead is not resolved in our modelling framework. To accurately estimate the diameter $\sigma$ of each bead, we compared the size of the globules found in the desorbed-collapsed (DC) phase in simulations to that of SAHF from fluorescence microscopy (see the paragraph Chromatin Bead Size below). We found that $\sigma \approx 70$ nm best matches the two sets of data.

As mentioned in the main text, we chose to model human chromosome 20 ($N = 6303$). Each bead was assigned to one of two possible types $q$: one for euchromatin (EC) ($q = 1$, coloured red in the figure) and the other for heterochromatin (HC) ($q = 2$, blue) (Figure~\ref{fig:model}A). Beads were coloured using chromatin immunoprecipitation with sequencing (ChIP-seq) data for H3K9me3~\citep{Chandra2012} and LaminB1~\citep{Sadaie2013} (see the subsection ChIP-seq Data Analysis below). We labelled beads as HC if the corresponding genomic region contains a peak in the H3K9me3 and/or LaminB1 data. Beads corresponding to the centromeric region ($26.4$ to $29.4$ Mb) were also marked as HC. All other beads were identified as EC. We note that recent work suggested that another process which can organise chromatin locally is loop extrusion by SMC complexes~\citep{Fudenberg2016}. For simplicity, we disregard this in our model, as we are interested in global chromatin organisation at the Mb level and above, whereas SMC/cohesin-mediated loops have a typical size of $\sim 100$ kb and are thought to be important for organisation at the sub-TAD and TAD level (for example, in experiments where cohesin is degraded in live cells, Hi-C data show a loss of TAD and loop structures, but larger scale features such as compartments persist~\citep{Schwarzer2017,Rao2017}).

The nuclear lamina (NL) can be modelled in at least two different ways. As done previously~\citep{Kinney2018}, one could represent the lamina by a smooth, attractive wall using a Lennard-Jones (LJ) potential; however, such a representation can only capture a coarse-grained interaction with the lamina. An alternative approach, implemented here, is to consider the lamina as a layer of beads ($q = 3$), which represents the lamins and other lamina-associated proteins that are part of the NL. This approach can more accurately account for the one-to-one interaction between NL proteins and chromatin. To generate the NL, we put 2000 lamina beads randomly within a $1\sigma \approx 70$ nm thick region just beneath the top of the simulation box (Figures~\ref{fig:model}A,B). We chose the lamina beads to be static in the simulation, as it is reasonable to believe that the dynamics of the NL constituents are much slower than that of chromatin.

To probe the structure of the chromosome, we performed molecular dynamics simulations with an implicit solvent (i.e., the nucleoplasm) using a scheme known as Brownian or Langevin dynamics. We simulated the chromosome in a cubic box with a linear dimension $L = 35\sigma \approx$ 2.5 $\mu$m. This size gives a volume fraction of chromatin of about 8\%. We employed periodic boundaries in the $x$ and $y$ direction but fixed boundaries in the $z$ direction due to the lamina wall. We used potentials common in polymer physics to simulate the chromatin fibre. First, a purely repulsive Week-Chandler-Andersen (WCA) potential was used to model steric interactions between beads
\begin{align}
\label{eqn:WCA}
U_{\text{WCA}}^{ij} = 4k_BT\left[ \left(\frac{\sigma}{r_{ij}}\right)^{12}\hspace{-0.2cm}-\left(\frac{\sigma}{r_{ij}}\right)^{6} + \frac{1}{4} \right] (\delta_{i+1,j} + \delta_{i-1,j})
\end{align}
if $r_{ij} < 2^{1/6}\sigma$ and 0 otherwise, where $r_{ij}$ is the separation between beads $i$ and $j$, and $\delta_{ij}$ is the Kroncker delta (i.e., $\delta_{ij} = 1$ if $i = j$ and $0$ otherwise). Second, a finite extensible non-linear elastic (FENE) spring acting between consecutive beads was used to enforce chain connectivity
\begin{align}
U_{\text{FENE}}^{ij} = - \frac{K_fR_0^2}{2}\ln\left[1-\left(\frac{r_{ij}}{R_0}\right)^2\right](\delta_{i+1,j} + \delta_{i-1,j}),
\end{align}
where $R_0 = 1.6\sigma$ is the maximum separation between the beads and $K_f = 30k_BT/\sigma^2$ is the spring constant. The superposition of the WCA and FENE potential with the chosen parameters gives a bond length which is approximately equal to $\sigma$~\citep{Brackley2013}. The interactions between non-consecutive beads were modelled using a truncated and shifted LJ potential
\begin{align}
U_{\text{LJ}}^{ij} = \frac{4\epsilon_{q_iq_j}}{\mathcal{N}} \left[\left(\frac{\sigma}{r_{ij}}\right)^{12} \hspace{-0.2cm}- \left(\frac{\sigma}{r_{ij}}\right)^{6} \hspace{-0.1cm}- \left(\frac{\sigma}{r_c^{q_iq_j}}\right)^{12} \hspace{-0.2cm}+ \left(\frac{\sigma}{r_c^{q_iq_j}}\right)^{6}\right] \left(1 - \delta_{i+1,j}\right)\left(1 - \delta_{i-1,j}\right)
\end{align}
if $r_{ij} \leq r_c^{q_iq_j}$ ($q_i$ is the type of bead $i$) and  $0$ otherwise, where $r_c^{q_iq_j}$ is the cutoff distance of the potential (set to $1.8\sigma$ for attractive interactions and $2^{1/6}\sigma$ otherwise) and $\mathcal{N}$ is a normalisation constant to ensure the depth of the potential is equal to the interaction energy $\epsilon_{q_iq_j}$. We set $\epsilon_{q_iq_j} = \epsilon_{\text{HH}}$ (in $k_BT$) for the interaction between HC beads ($q_i,q_j=2$), $\epsilon_{q_iq_j} = \epsilon_{\text{EE}}$ for that between EC beads ($q_i,q_j=1$), and $\epsilon_{q_iq_j} = \epsilon_{\text{HL}}$ for that between HC and NL beads ($q_i,q_j = 2$ or $3$, $q_i\neq q_j$). Other interactions were purely repulsive ($\epsilon_{q_iq_j} = 1$). There is no interaction between lamina beads as they are static in the simulation. The sum of potential energy terms involving bead $i$ is
\begin{align}
U_i = \sum_{j\neq i} \left(U_{\text{WCA}}^{ij} + U_{\text{FENE}}^{ij} + U_{\text{KP}}^{ij} + U_{\text{LJ}}^{ij}\right).
\end{align}
The time evolution of each bead along the fibre is governed by the following Langevin equation
\begin{align}
\label{eqn:EOM}
m_i\frac{d^2\vec{r}_i}{dt^2} = - \nabla U_i - \gamma_i \frac{d\vec{r}_i}{dt} + \sqrt{2k_BT\gamma_i}\vec{\eta}_i(t),
\end{align}
where $m_i$ and $\gamma_i$ are the mass and the friction coefficient of bead $i$, and $\vec{\eta}_i$ is its stochastic noise vector with the following mean and variance
\begin{align}
\langle\bm{\eta}(t)\rangle = 0;\quad \langle\eta_{i,\alpha}(t)\eta_{j,\beta}(t')\rangle = \delta_{ij}\delta_{\alpha\beta}\delta(t-t'),
\end{align}
where the Latin and Greek indices run over particles and Cartesian components, respectively, and $\delta(t-t')$ indicates the Dirac delta function. The last term of Eq.~(\ref{eqn:EOM}) represents the random collisions caused by the solvent particles. We assumed all beads have the same mass and friction coefficient (i.e., $m_i = m$ and $\gamma_i = \gamma $) and set $m = \gamma = k_BT = 1$. We used the Large-scale Atomic/Molecular Massively Parallel Simulator (LAMMPS) (\url{http://lammps.sandia.gov})~\citep{Plimpton1995} to numerically integrate the equations of motion using the standard velocity-Verlet algorithm. For the simulation to be efficient yet numerically stable, we set the integration time step to be $\Delta t = 0.01\tau_{\text{Br}}$, where $\tau_{\text{Br}}$ is the Brownian time, or the typical time for a bead to diffuse a distance of its own size (i.e., $\tau_{\text{Br}} = \sigma^2/D$ with $D$ being the diffusion coefficient).

\paragraph{Initial Conditions and Equilibration}
We initialised the chromatin fibre as an ideal random walk, in a larger box ($L = 100\sigma$, fixed boundaries) in which the lamina is absent. We allowed the fibre to equilibrate for $10^4\tau_{\text{Br}}$, during which the beads can only interact via steric repulsion (with chain connectivity maintained). We used the soft potential in the first $6\times 10^3\tau_{\text{Br}}$ to remove overlaps in the polymer such that it becomes a self-avoiding chain. In formula, this potential is given by
\begin{align}
U_{\text{Soft}}^{ij} = A\left[1+\cos\left(\frac{\pi r_{ij}}{r_c}\right)\right]
\end{align}
if $r_{ij} < r_c$ and 0 otherwise, where $r_c = 2^{1/6}\sigma$ and $A$, the maximum of the potential, gradually increases from $0$ to $100k_BT$. We reverted to the WCA potential (for interaction between all beads) for the remaining part of this equilibration period. In the next $5\times 10^3\tau_{\text{Br}}$, we compressed the simulation box incrementally to the desired volume ($L = 35\sigma$) using indented walls. We then generated the lamina beads at the top of the simulation box as described above.  Finally, we let the chromatin fibre equilibrate with the lamina (interacting via steric repulsion) for $5\times 10^3\tau_{\text{Br}}$, with boundary conditions identical to those for the main simulation run. 

\paragraph{Observables}
As discussed briefly in the main text, we considered two observables to quantify the state of the system in the parameter space ($\epsilon_{\text{HH}},\epsilon_{\text{HL}}$). First, to determine whether the system is adsorbed or desorbed, we measured the distance $\bar{z}$ between the centre of mass of the chromatin fibre and the NL. $\bar{z}$ can be expressed as
\begin{align}
\bar{z} = \frac{1}{N}\sum_{i=1}^{N}z_i,
\end{align}   
where $z_i$ is the distance of bead $i$ from the NL (note that all beads have the same mass). Second, to determine whether the system has an extended or collapsed polymer conformation, we measured the average local number density $\rho$ of neighbouring beads. $\rho$ is defined as
\begin{align}
\rho = \frac{1}{4/3\pi N r_s^3} \sum_{i=1}^{N}\sum_{j=1}^{N}\Theta(r_s - r_{ij}),
\end{align}
where $\Theta(x) = 1$ if $x \geq 0$ and $0$ otherwise. In essence, this observable counts the number of beads within a sphere of radius $r_s$ around each bead $i$. The threshold $r_s$ was set to $5\sigma \approx 350$ nm. We have verified that choosing other biophysically reasonable values for $r_s$ (e.g., $3\sigma$ and $7\sigma$) does not much affect the phase diagram of the system. In particular, the transition line separating the extended and collapsed regime obtained from $\rho$ remains consistent with that obtained from $\bar{z}$ (see the cyan lines in Figures~\ref{fig:phase_diagram}A,B).

\paragraph{Chromatin Bead Size}
In this work, we reported distances in units of the size/diameter $\sigma$ of each chromatin bead. To estimate the size of each bead, we compared the size of the heterochromatin (HC) globules found in the desorbed-collapsed (DC)/senescence state in our simulations to that of senescence-associated heterochromatin foci (SAHF) identified in fluorescence microscopy. In our simulations, we calculated the radial distribution profile of HC beads in the globule (see Figure~\ref{fig:phase_diagram}G). The effective radius of the globule was then identified by the inflection point of the profile. Averaging over 10 simulations (with $\epsilon_{\text{HH}} = 1.4$ and $\epsilon_{\text{HL}} = 0.2$), we found the mean radius of the globule to be $r_{\text{DC}} = 9.21\pm0.09\,\sigma$. We estimated the size of SAHF based on the image with H3K9me3 staining (see Figure~\ref{fig:phase_diagram}H). We computed the area $A_{\text{SAHF}}$ of the SAHF in focus using Fiji distribution of ImageJ~\citep{Schindelin2012}. The effective radius of the SAHF was estimated as $r_{\text{SAHF}} = (A_{\text{SAHF}}/\pi)^{1/2}$. From averaging over 12 SAHF, we found a mean radius $r_{\text{SAHF}} = 0.64\pm0.04$ $\mu$m. Setting $r_{\text{DC}} = r_{\text{SAHF}}$, this gives $\sigma = 70\pm5$ nm.

It is possible to verify that this estimation is within the right order of magnitude from calculating a lower and upper bound estimate of the bead size. Since we have chosen each bead to contain 10 kb of chromatin, there are roughly $N_{\text{nuc}} = 50$ nucleosomes in each bead (as the number of bp per nucleosome is around 200). Approximating each nucleosome as a cylinder with radius $r \approx 5$ nm and height $h = r \approx 5$ nm, its volume is therefore $V_{\text{nuc}} \approx 125\pi$ $\text{nm}^3$. The minimum bead size $\sigma_{\text{min}}$ is achieved by packing the nucleosomes tightly in the bead with no empty space. Setting the volume of the bead equal to the total volume of all nucleosomes, the effective diameter of the bead is
\begin{align}
\sigma_{\text{min}} = \left(\frac{6}{\pi}N_{\text{nuc}}V_{\text{nuc}}\right)^{1/3} \approx 33 \text{ nm}. 
\end{align}
To obtain a high-end estimate of the bead size, we need to make assumptions about the structure of chromatin within the bead. Specifically, there are three estimates required: (i) the linear compaction of chromatin (i.e., the number of nucleosomes per nm, or typically reported in nuc/11 nm), (ii) the Kuhn length (twice the persistence length) of the chromatin fibre and (iii) its conformation within the bead. The linear compaction of chromatin remains poorly known. We here take a conservative estimate of 2 nuc/11 nm reported by~\citet{Dekker2008} from 3C experiments on yeast chromosomes. With 50 nucleosomes in a bead, the contour length of the chromatin fibre packed in a bead is $L \approx 275$ nm. For the Kuhn length $b$ of chromatin, Hi-C cyclization experiments have suggested that this is around 1 kb for mammalian chromatin fibres~\citep{Sanborn2015}. Using the linear compaction ratio above, we find $b \approx 27.5$ nm. This value is compatible with the range of Kuhn lengths estimated for \textit{Drosophila} chromatin~\citep{Lesage2019}. For the conformation of the chromatin fibre, we approximate the fibre to behave like a self-avoiding chain to obtain the largest estimate of the bead size. The radius of gyration $R_g$ for a real chain is related to its contour length $L$ and its Kuhn length $b$ roughly by $R_g \approx b(L/b)^\nu/\sqrt{6}$, where $\nu \approx 0.588$~\citep{Rubinstein2003}. Putting everything together, a high-end estimate of the bead size is
\begin{align}
\sigma_{\text{max}} = 2R_g \approx \frac{2b}{\sqrt{6}}\left(\frac{L}{b}\right)^{0.588} \approx 87 \text{ nm}. 
\end{align}
It is evident that our estimate of the bead size based on experimental data (i.e., $\sigma \approx 70$ nm) lies within these two bounds, suggesting that it is appropriate and realistic.

\subsubsection{Contact Maps from Simulations and Hi-C Experiments}
The contact maps for the OCI values in Figure~\ref{fig:oci} were obtained as follows. In our simulations, we generated Hi-C-like contact maps by calculating the probability of two beads being in contact, i.e., their separation distance is closer than $3\sigma$. This probability was determined from computing the frequency of beads in contact over a $5\times 10^4\tau_{\text{Br}}$ time period in each simulation run, which was then averaged over the 20 simulations performed for each cell state. For the progeria Hi-C experiment~\citep{Mccord2013}, we constructed contact maps (200 kb resolution) based on the valid pairs reported in that reference (see Key Resources Table). In particular, we used the Age Control sample for growing and HGPS-p19 sample for progeria. For the senescence experiment~\citep{Chandra2015}, we obtained the raw sequencing data for both growing and senescent cells. We used the HiC-Pro pipeline (version 2.10.0)~\citep{Servant2015} to process the sequencing data (aligned to the Ensembl GRCh37 human genome) and generate the contact maps (50 kb resolution). All experimental contact maps were normalised using the standard iterative correction procedure~\citep{Imakaev2012} to eliminate experimental biases. 

\subsubsection{Open Chromatin Index (OCI)}
When comparing the contact maps, we considered the open chromatin index (OCI), which is a ratio of the distal contact strength to the local contact strength. More precisely, we defined the (normalised) local contact signal $\mathcal{C}_\ell$ and distal contact signal $\mathcal{C}_d$ for each chromatin bin (say bin $i$) as:
\begin{align}
\mathcal{C}_\ell(i) &= \frac{1}{N_\ell(i)}\sum_{j=1}^N c_{ij}\Theta(s_d-s_{ij}) \\
\mathcal{C}_d(i) &= \frac{1}{N_d(i)}\sum_{j=1}^N c_{ij}\Theta(s_{ij}-s_d),
\end{align}
where $c_{ij}$ is the contact probability between chromatin segments in bins $i$ and $j$, $s_{ij}$ is their genomic separation, $N_\ell(i)$ and $N_d(i)$ are the number of possible local and distal contact pairs, respectively, for bin $i$, and $s_d$ is the genomic distance beyond which we consider contacts to be distal. We set $s_d = 2$ Mb as this is close to the maximum size of a TAD~\citep{Dekker2015}, so contacts beyond this distance are considered to be non-local. We then used these quantities to define the Open Chromatin Index (OCI) as
\begin{align}
\mathrm{OCI}(i) &= \log_2\left(\frac{\mathcal{C}_d(i)}{\mathcal{C}_\ell(i)}\right).
\end{align}
Note that the OCI is not well-defined in genomic regions that are difficult to be sequenced and have no contact signals, such as the centromeric region.

\subsection{Quantification and Statistical Analysis}
\subsubsection{Mass Spectrometry Quantification}
Proteome Discoverer v1.4 (Thermo Fisher Scientific) and Mascot (Matrix Science) v2.2 were used to process raw mass spectrometry data files. Data were aligned with the UniProt human database, in addition to using the common repository of adventitious proteins (cRAP) v1.0. Protein identification allowed an MS tolerance of $\pm$20 ppm and an MS/MS tolerance of $\pm$0.1 Da along with up to 2 missed tryptic cleavages. Quantification was achieved by calculating the sum of centroided reporter ions within a $\pm$2 millimass unit (mmu) window around the expected $m/z$ for each of the 8 TMT reporter ions.

\subsubsection{ChIP-seq Data Analysis}
As our simulation model input, ChIP-seq data for H3K9me3 and LaminB1 were downloaded from NCBI Geo repositories GSE38448 and GSE49341, respectively.  Reads were assessed for quality with FastQC, and poor quality ends (phred-scores $<$ 30) were filtered using the Trim Galore package.  High-quality sequences were subsequently aligned to the Ensembl GRCh37 human genome with bowtie2, before consecutively removing non-unique alignments and repetitive sequences with SAMtools and Picard tools (version 1.98).  Regions of H3K9me3 or LaminB1 occupancy were determined using the RSEG package (mode 2) against the non-specific, paired input controls with standard settings.  Where multiple replicates existed, BEDtools was used to intersect peaks and generate a set on consensus regions that exist in the majority of replicates.

\subsubsection{Statistical Analysis}
Statistical tests performed in this work include the two-sample Kolmogorov-Smirnov (KS) test for $\Delta$OCI and the Pearson correlation score for the OCI values between simulations and experiments.  These test statistics were calculated using the Scipy package in Python. The number of simulations performed for the results reported in each figure is stated in the caption. Results were deemed to be significant when $p < 0.05$, which was the case for the statistical tests conducted.  

\subsection{Data and Code Availability}
The mass spectrometry data generated during this study are available at the PRIDE archive (PXD014929). Original/source data for the figures in the paper are available at Edinburgh DataShare (\url{https://doi.org/10.7488/ds/2593}).

\section{Supplemental Items}
\begin{itemize}
\item {\footnotesize{\textbf{Table S1: Mass spectrometry data showing the abundance of individual proteins following chromatin fractionation in growing cells and in senescent cells}, Related to Figure~\ref{fig:phase_diagram} and Figure S1.}}
\end{itemize}

\nocite{Chandra2012,Sadaie2013,Chandra2015,Mccord2013,Jeanblanc2012,Young2009,Quinlan2010,Langmead2012,Servant2015,Schindelin2012,Plimpton1995,Song2011,Li2009}


\setcounter{figure}{0}
\renewcommand{\figurename}{\textbf{Figure S}}
\renewcommand{\thefigure}{\textbf{\arabic{figure}}}

\section{Supplemental Figures}
\begin{figure*}[h]
  \centering
  \includegraphics[width=174mm]{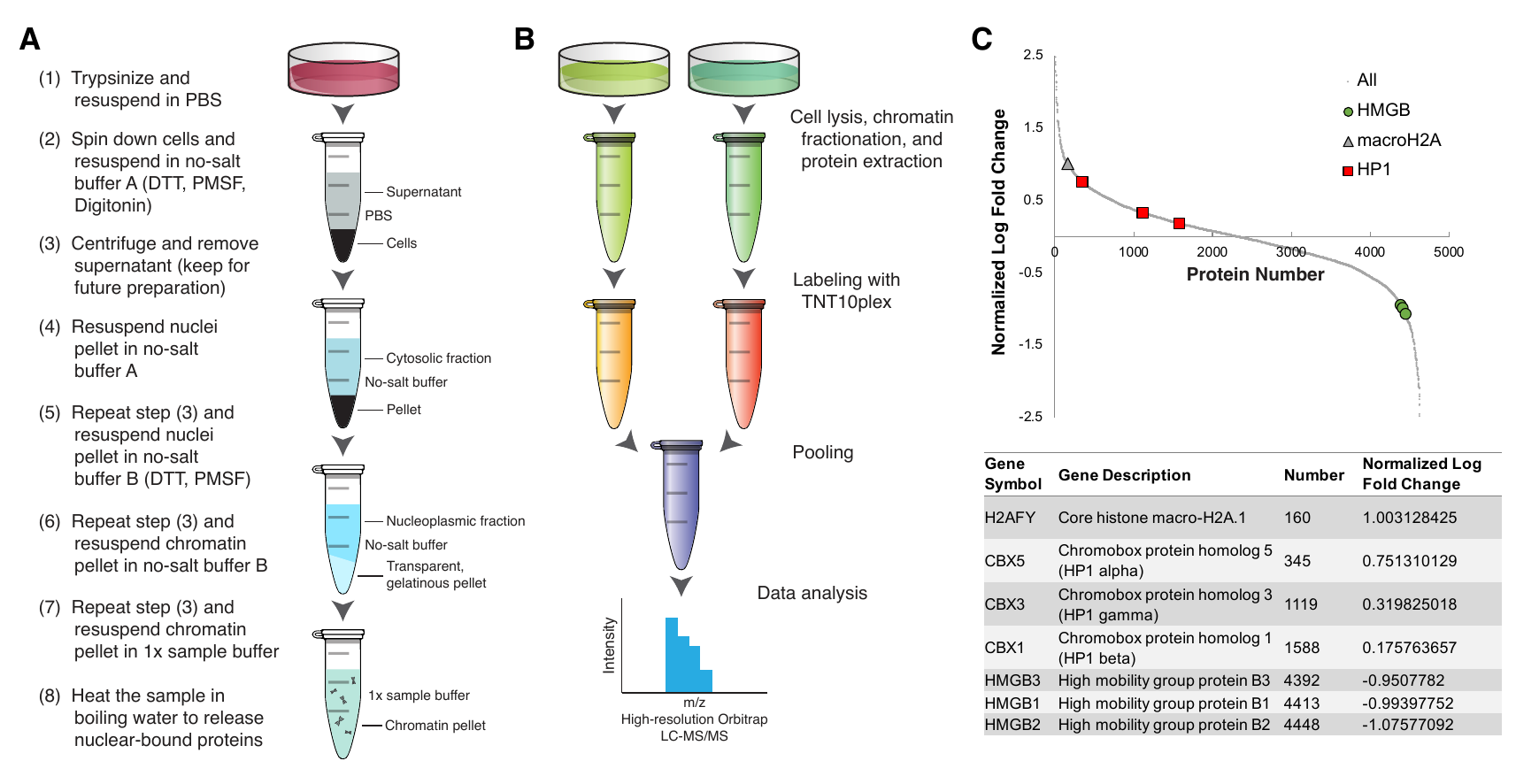}
  \caption{{\bf{Chromatin fractionation and mass spectrometry experiments investigating the change in protein abundance between growing and senescent cells}}, related to Figure~\ref{fig:phase_diagram} and STAR Methods. (A) The key stages for chromatin isolation of a single sample. (B) A schematic representation of chromatin fractionation from different samples followed by identification and quantitation of proteins based on high-resolution Orbitrap LC-MS/MS. (C) \textit{Top}: A graph showing the normalized log fold change of individual protein abundance between growing and senescent cells. Proteins are numbered from the highest to the lowest ratio. \textit{Bottom}: A table listing the normalized log fold change for macroH2A, HP1, and high mobility group proteins.}
  \label{fig:mass_spec}
\end{figure*}

\begin{figure*}[h]
  \centering
  \includegraphics[width=134mm]{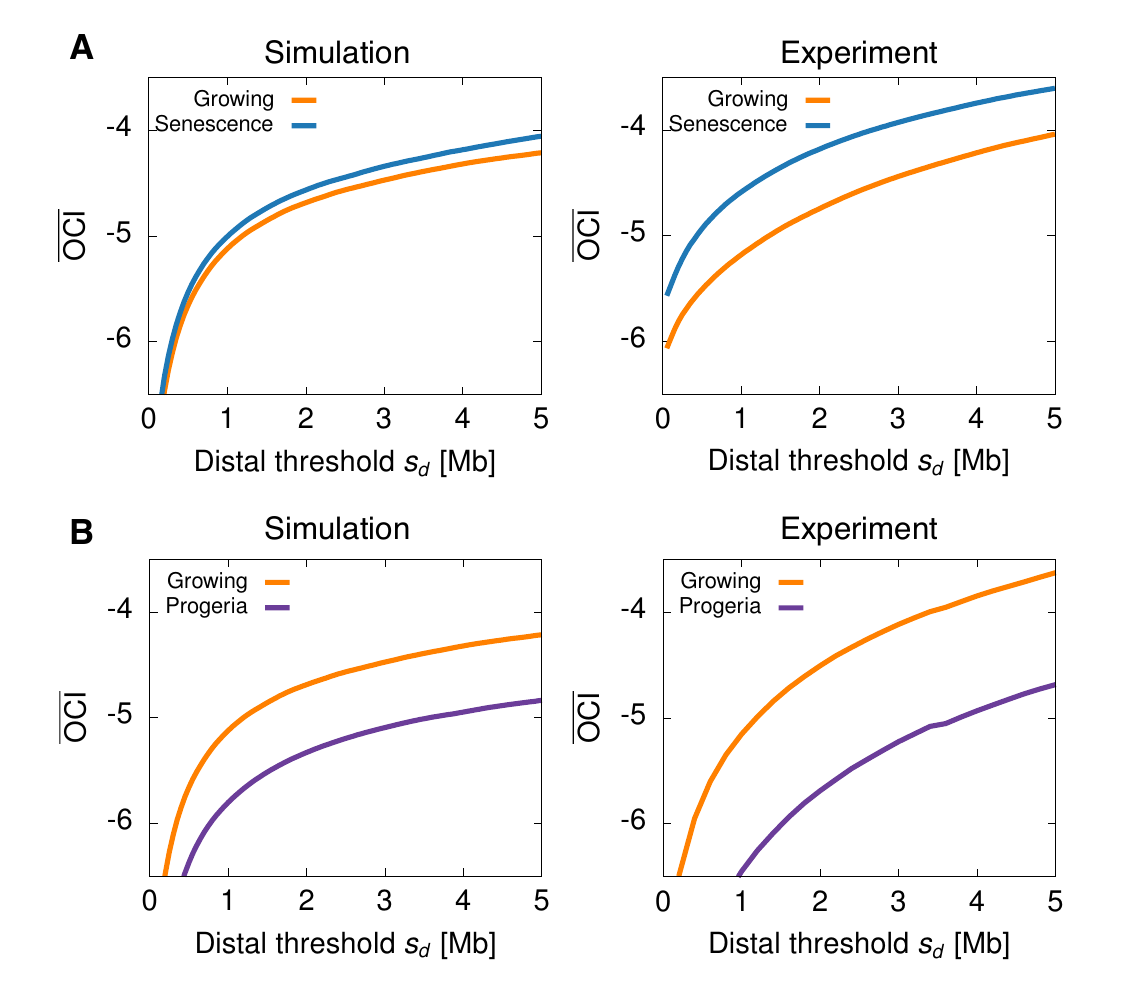}
  \caption{{\bf{Varying the threshold $s_d$ which distinguishes local from distal contacts does not alter the general trends for the change in the open chromatin index (OCI) between different cell states}}, related to Figure~\ref{fig:oci}. (A and B) The plots compare the average OCI value between (A) growing and senescent state and (B) growing and progeroid state for different thresholds $s_d$. Note that the OCI increases from growing to senescence and decreases from growing to progeria for different $s_d$, both in simulations and in experiments.}
  \label{fig:oci_threshold}
\end{figure*}

\begin{figure*}[h]
  \centering
  \includegraphics[width=94mm]{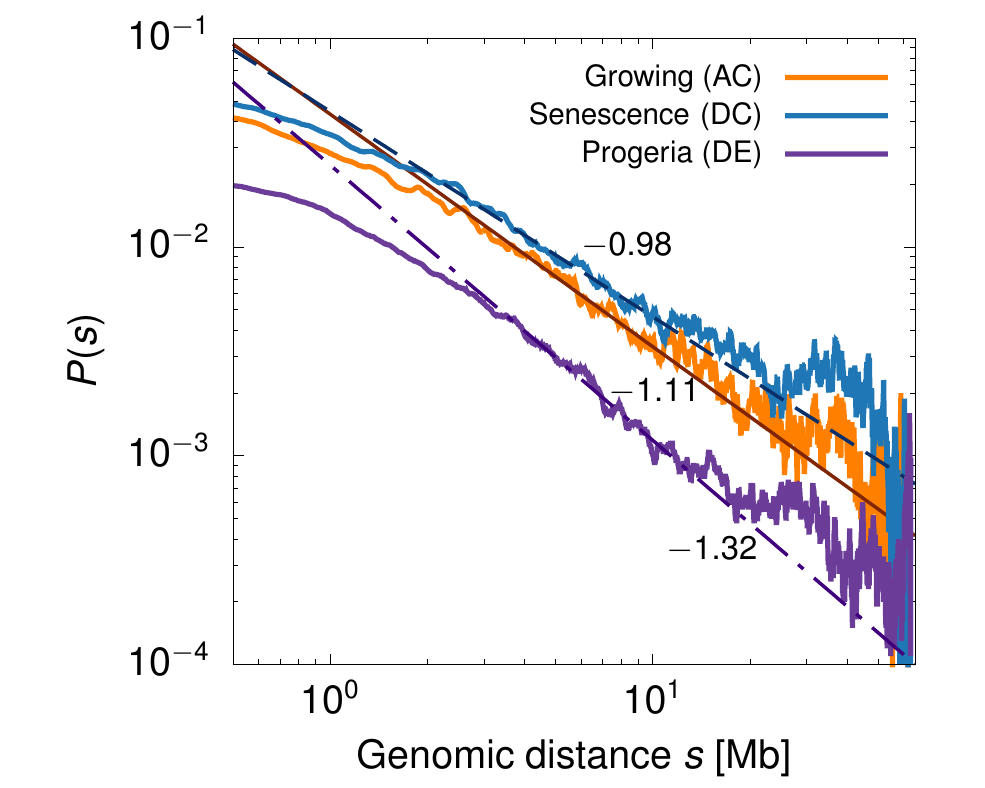}
  \caption{{\bf{Contact probability $P(s)$ as a function of genomic distance $s$}}, related to Figure~\ref{fig:oci}. A log-log plot showing the decay in the contact probability $P(s)$ as the genomic distance $s$ between two chromatin segments increases, for growing (adsorbed-collapsed, AC), senescence (desorbed-collapsed, DC), and progeria (desorbed-extended, DE). Straight lines are linear fits to the log-log curves with the measured exponents $c$ shown for the different cell states.}
  \label{fig:prob_decay}
\end{figure*}

\end{document}